\documentclass[lettersize,journal]{IEEEtran}
\AtBeginDocument{%
  }

\usepackage{multirow}
\usepackage{subcaption}
\usepackage{xspace}
\usepackage{xcolor}
\usepackage{graphicx}
\usepackage{booktabs}
\usepackage{makecell}
\usepackage{array}
\newcolumntype{C}[1]{>{\centering\arraybackslash}m{#1}}
\usepackage{comment}
\usepackage{xcolor}
\usepackage{arydshln}
\usepackage{algpseudocode}
\usepackage{url}
\usepackage[table]{xcolor}  % Adds table support for arrayrulecolor
\definecolor{codebg}{gray}{0.95}
\begin{document}

%%
%% The "title" command has an optional parameter,
%% allowing the author to define a "short title" to be used in page headers.
% \title{Accelerating Topology Optimization on AMD Versal AIE-ML Engines}
% Title for Arxiv
\title{Accelerating CRONet on AMD Versal AIE-ML Engines}

\author{
Kaustubh Manohar Mhatre$^{1}$, 
Vedant Tewari$^{1}$, 
Aditya Ray$^{1}$, 
Farhan Khan$^{2}$,\\
Ridwan Olabiyi $^{1}$,
Ashif Iquebal $^{1}$,
Aman Arora$^{1}$ \\
$^{1}$Arizona State University, Tempe, USA, 
$^{2}$Technical University of Munich, Germany \\
\{kmhatre, vtewari1, adityar4, rolabiyi, aiquebal, aman.kbm\}@asu.edu, 
farhan.khan@tum.de}

%%
%% The abstract is a short summary of the work to be presented in the
%% article.
\newcommand{\versal}{AMD Versal\xspace}
\newcommand{\red}{\textcolor{black}}
\newcommand{\rev}{\textcolor{black}}

%%
%% This command processes the author and affiliation and title
%% information and builds the first part of the formatted document.
\maketitle
\begin{abstract}
Topology optimization is a computational method used to determine the optimal material distribution within a prescribed design domain, aiming to minimize structural weight while satisfying load and boundary conditions. 
For critical infrastructure applications, such as  structural health monitoring of bridges and buildings, particularly in digital twin contexts, 
% achieving real-time 
\rev{low-latency energy-efficient}
topology optimization is essential. 
Traditionally, topology optimization relies on finite element analysis (FEA), a computationally intensive process. 
Recent advances in deep neural networks (DNNs) have introduced data driven alternatives to FEA, substantially reducing computation time while maintaining solution quality.  
%However, the inference latency of these neural network models, especially for complex designs, still remains a major obstacle to real-time deployment.
These DNNs have complex architectures and implementing them on inference-class GPUs results in high latency and poor energy efficiency. 
To address this challenge, we present a hardware accelerated implementation of a topology optimization neural network (CRONet) on the AMD Versal AI Engine-ML (AIE-ML) architecture. 
Our approach efficiently exploits the parallelism and memory hierarchy of AIE-ML engines to optimize the execution of various neural network operators. 
\rev{We are the first to implement an end-to-end neural network fully realized on the AIE-ML array, where all intermediate activations and network weights reside on-chip throughout inference, eliminating any reliance on 
% the PL fabric or 
DRAM for intermediate data movement.
% We also present a library of parameterizable, high-performance operator kernels implemented on the Versal AIE-ML array, designed to be scalable and adaptable to support other neural networks.
Experimental results demonstrate that our implementation achieves up to 2.49x improvement in latency and up to 4.18x improvement in energy efficiency compared to an inference-class ML-optimized GPU in the same power budget (Nvidia T4) after scaling for technology node.}
These results highlight the potential of Versal AIE-ML based acceleration for enabling low-latency energy-efficient topology optimization.
\end{abstract}

%%%%%%%%THOUGHTS/NOTES
%WHy not use the PL? Why Versal? -> We can say two things: 1. this leaves the PL free for other things. 2. this means no PL-AIE transfers, everything is on the AIE, so better latency. Kind of a weak argument, but that's the best we can say for now. It'd be better to have a AIE+PL implementation and compare with that, or just choose an AIE-only device, or find some pre-/post-processing to do on the PL.

%We have to say somewhere why we only used GMIO and not PLIO -> We need to say that since everything happens on the AIEs, the bandwidth requirement is very low. Hence, we don't need to use PLIO

%Why do you need to run this real-time? What's the real-time requirement? -> I think we should not focus on this. FOcus on just low latency and energy-efficiency

%Compare to an FPGA implementation -> We don't have reuslts, but we can talk in terms of programmability.

%We need to say that this is a very complex network with very unique kernels and unique dataflow (CNN+RNN). It has a uniqueness that dependency not very efficient for pipelining and can't do batching. -> May be we add a section for characterization of the DNN workload. WHile we do this, we have to talk about how accuracy is measured and why BF16 is good (this is a point that an FCCM reviewer had asked)

\section{Introduction}
Topology Optimization (TO) is a powerful technique used to identify the most efficient material distribution within a defined geometric design domain, aiming to reduce weight while meeting specified load requirements.
It has become an essential tool across various modern engineering fields, including aerospace, automotive, and civil engineering, facilitating the design of lightweight and durable structures. 
Traditional TO workflows are bottlenecked by the iterative nature of Finite Element Analysis (FEA), which requires significant simulation time and substantial hardware resources \cite{deaton:2014:to_review}. 
Consequently, its use has been largely restricted to offline design applications where 
% real-time performance 
\rev{low latency and energy efficiency}
are not a priority.

However, the scope of TO has expanded significantly. Beyond design, it plays an increasingly important role in monitoring existing structures, such as bridges and buildings, and performing structural health assessments to ensure their long-term reliability \cite{catbas:2008:shm_to}. 
These applications are particularly critical in the context of digital twins, where the continuous tracking and evaluation of physical structures \rev{requires low latency.} 
To bridge this gap, researchers have recently turned to neural network-based approximations \cite{olabiyi:2025:cronet} to replace the intensive computations of FEA.
A prominent example is CRONet \cite{olabiyi:2025:cronet}, a high-speed surrogate model designed to approximate FEA results by leveraging a hybrid architecture of Convolutional Neural Networks (CNNs) and Recurrent Neural Networks (RNNs).
While deep learning models can significantly reduce the time required for a TO cycle, standard hardware deployments still struggle to meet the low latency and energy efficiency demands for digital twin integration.

% These neural networks present a new challenge, as these networks are typically large and computationally intensive.
\rev{The hybrid nature of CRONet, with a complex set of operators such as Convolution 3D, sigmoid linear unit activation (SILU) and 3D adaptive average pooling, makes it challenging to accelerate. Furthermore, the iterative process of TO does not allow batching, as a generated output becomes the input for the next iteration.
}
%, as these networks are typically large and computationally intensive.
Thus, while GPUs are the standard for neural network training, they are suboptimal for low-latency and energy-efficient inference, especially for TO applications. 
%CPUs, designed for general-purpose serial processing, exhibit very high latency for neural network execution.
%Conversely, while GPUs offer high parallel throughput, they struggle in low-latency \cite{Jouppi:2017:sigarch:tpu} use cases and consume substantial power.
While FPGAs offer a highly configurable computing substrate, they remain challenging to program and exhibit relatively low computational throughput per unit area.
% The AMD Versal AI Engine-ML (AIE-ML) architecture is an attractive target for this acceleration, as its highly parallel, heterogeneous cores can be configured in a dataflow manner, making it ideal for low-latency neural network execution.
\rev{The AMD Versal AI Engine-ML (AIE-ML) architecture \cite{amd:web:versal} is an attractive target for this acceleration, as its highly parallel, heterogeneous cores can be configured in a dataflow manner, making it well-suited for low-latency neural network execution. 
The overall Versal architecture comprises the Processing System (PS), Programmable Logic (PL), and the AIE-ML array.
Within this architecture, the AIE-ML array \cite{amd:2023:web:aie2_arch} consists of a grid of vector processors alongside memory storage elements called Memory Tiles. The vector processors provide high compute density, while the memory tiles offer higher storage capacity and support efficient data manipulation.}
Although existing work on neural network acceleration for Versal \cite{zhuang:2023:fpga:charm,zhuang:2023:fpga:automm,mhatre:2024:gama,taka:2023:fpt:maxeva} has focused on accelerating individual computational layers, such as matrix multiplication, 
% there has been a limited focus on optimizing and accelerating the complete network architecture for AIE-ML.
\rev{to the best of our knowledge, no prior work has demonstrated running a complete neural network entirely on the AIE-ML array.}

\rev{
We present the first implementation of a hybrid CNN–RNN for TO accelerated on AMD Versal AIE-ML.
This is enabled by a fully on-AIE array inference pipeline that uses  persistent on-chip weights 
% PL involvement 
and minimizes off-chip memory traffic through Memory Tile buffering and fusion. 
To support this design, we develop custom kernels for each network layer and enhance the state-of-the-art GEMM mapping techniques.
We further introduce a congestion-aware placement strategy that enables scalable compilation by aligning kernel placement with dataflow locality, avoiding routing bottlenecks. 
Together, these contributions deliver up to 2.49× speedup over an NVIDIA T4 GPU and up to 4.2× higher energy efficiency, demonstrating the effectiveness of the Versal AIE-ML platform for low-latency, energy-efficient inference.
% We present a low-latency acceleration of a hybrid neural network .
% We also create parameterizable operators that incorporate optimizations such as layer fusion and on-chip weights to achieve true real-time performance at the edge.
}

Our contributions in this work are:
\rev{
\begin{itemize}
  \item \textbf{Hybrid Network Acceleration:} We are the first to map a hybrid CNN-RNN neural network model for Topology Optimization on the AMD Versal AIE-ML array.
  \item \textbf{Fully on-AIE DNN inference pipeline}: We map the complete DNN exclusively onto the AIE-ML array, retaining all weights and intermediate activations on-chip by leveraging AIE memory and Memory Tiles as staging buffers.
  % , thereby eliminating the latency and bandwidth overhead of repeated DRAM access.
  This reduces DRAM access to reading inputs and writing outputs only, enabling a streamlined dataflow implementation that achieves low-latency DNN inference.
  \item \textbf{Custom congestion aware placement:} 
  We develop a custom congestion-aware placement strategy that overcomes routing congestion in large-scale AIE designs, achieving high utilization and reliable compilation while significantly reducing compilation time.
  % We develop a custom congestion-aware placement strategy necessitated by the compiler’s inability to successfully map large designs due to routing congestion. Our approach enables high AIE utilization and ensures successful compilation, while significantly reducing overall compilation time.
  % We develop a custom congestion-aware placement algorithm for mapping kernels to specific AIE tiles and shared buffers to Memory Tiles. The compiler's automatic placement consistently fails for large designs due to routing congestion, making manual placement necessary to enable compilation at the scale required by this work. Our placement strategy clusters AIEs belonging to the same layer and maps them onto the array in a manner that accounts for data flow direction, ensuring that the placement of each tile is informed by the location of its input source and output destination. This achieves congestion-free routing and successful compilation of the full design, while also significantly reducing overall compilation time.
  \item \textbf{Performance Benchmarking:} Our implementation achieves up to 2.49 $\times$ speedup over 
  % the NVIDIA T4, 
  an inference-class ML-optimized GPU in the same power budget (Nvidia T4) after scaling for technology node while delivering up to 4.2 $\times$ superior energy efficiency, demonstrating the suitability of the Versal AIE-ML platform for low-latency, power-efficient inference workloads.
  \item \textbf{Custom parameterizable operators library:} 
  We develop highly parameterizable and optimized kernels for GEMM (General Matrix Multiply), 2D/3D convolution, max pooling, and 2D/3D adaptive average pooling, targeting the AIE-ML architecture. RNN layers are additionally supported through full unrolling. Our parameterization controls scaling not only within a single AIE but also across multiple AIEs in the array, enabling layers that exceed the memory capacity of a single AIE. The resulting parameterized operators can be reused directly or adapted with minor modifications to implement other network architectures.
  % We develop a  highly parameterizable and optimized kernels for 2D/3D convolution, max pooling, and 2D/3D adaptive average pooling specifically for the AIE-ML architecture. We also support RNNs with full unrolling. Our parameterization controls the kernel scaling not only for a single AIE but also across the AIE array. We call this subgraph parameterization. This enables support for a larger layer that cannot fit into a single AIE. Our parameterized operators (subgraph and kernel) can be easily used directly or with minor modifications to implement other networks.
  Our CRONet accelerator and operator implementations are publicly available at: https://github.com/xxxx (blinded for review)
\end{itemize}
}

\section{Background}\label{sec:background}

\subsection{CRONet: A Deep Learning Approach to Topology Optimization}

% Topology Optimization (TO) is a mathematical framework used to determine the optimal distribution of material within a defined design space, typically aimed at minimizing structural weight while satisfying specific mechanical constraints. 
% Traditionally, TO relies on Finite Element Analysis (FEA) to evaluate structural responses (such as displacement and stress) at every design iteration. 
% However, FEA is inherently sequential and computationally intensive, often requiring hundreds of iterations to reach convergence, which creates a significant bottleneck for real-time applications.

\begin{figure}[t]
  \centering
  \includegraphics[width=1\linewidth]{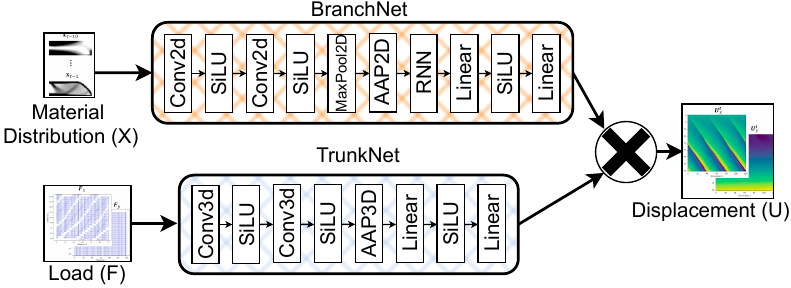}
  \caption{Architecture of CRONet \cite{olabiyi:2025:cronet}.}
  \label{fig:cronet_arch}
  \vspace{-5mm}
\end{figure}

Ridwan et al. \cite{olabiyi:2025:cronet} introduced Convolutional Recurrent Operator Approximator Network (CRONet), a high-speed surrogate model that approximates the FEA solver within the optimization loop, significantly reducing latency without compromising solution accuracy.

% \textbf{CRONet Architecture and Operation}
\rev{CRONet belongs to the DeepONet \cite{lu:2021:deeponet} family of neural operators and employs a dual-network architecture, as shown in Figure \ref{fig:cronet_arch}. 
Unlike standard models, CRONet exhibits sequential dependencies where the output of each optimization iteration serves as input to the next, preventing parallelization across iterations. 
This recurrent structure, combined with the interplay between its CNN and RNN components, makes efficient hardware mapping a non-trivial challenge.}
% CRONet is based on the DeepONet \cite{lu:2021:deeponet} family of neural operators.
% Figure \ref{fig:cronet_arch} shows the architecture of CRONet.
% It utilizes a dual-network architecture to process different aspects of the physical problem:

The dual-network architecture consists of:
(1) \textbf{BranchNet:} 
    % This network processes the evolving material distribution ($X$). Unlike standard surrogate models, the Branch Network employs a Convolutional Recurrent Neural Network (CRNN). 
    % This hybrid structure is critical as it captures both the spatial features of the material layout and the temporal dependencies across successive optimization iterations.
    The Branch Network processes the evolving material distribution ($X$) using a Convolutional Recurrent Neural Network (CRNN), distinguishing it from standard surrogate models. 
    This hybrid structure captures both the spatial features of the material layout through its convolutional layers and the temporal dependencies across successive optimization iterations through its recurrent layers.
(2) \textbf{TrunkNet:} The Trunk Network processes the fixed load configuration ($F$). Since the loading conditions typically remain constant throughout the optimization process, it employs a standard Convolutional Neural Network (CNN) to extract the spatial features of the applied forces.
    % This network processes the fixed load configuration ($F$). Since the loading conditions typically remain constant throughout the optimization process, a standard Convolutional Neural Network (CNN) is used to extract the spatial features of the applied forces.
% \end{itemize}

The outputs of the BranchNet and TrunkNet are combined to produce the final displacement field ($U$), representing the predicted structural response.
In a typical CRONet enhanced workflow, the optimization begins with a limited number of ``warm-up" iterations (e.g., the first 10 iterations) performed using a traditional FEA solver. 
The resulting solutions provide the initial context for the recurrent layers. 
CRONet then replaces the FEA solver for subsequent iterations, providing near-instantaneous displacement field predictions and sensitivity analysis. 
By bypassing repetitive and costly FEA computations, this hybrid approach has demonstrated speedups of up to 78\% on benchmark problems such as the Messerschmitt-Bölkow-Blohm (MBB) beam, while maintaining design compliance within 5\% of traditional FEA solutions \cite{olabiyi:2025:cronet}.

\subsection{Versal for Hardware Acceleration}

\begin{figure}[t]
  \centering
  \includegraphics[width=1\linewidth]{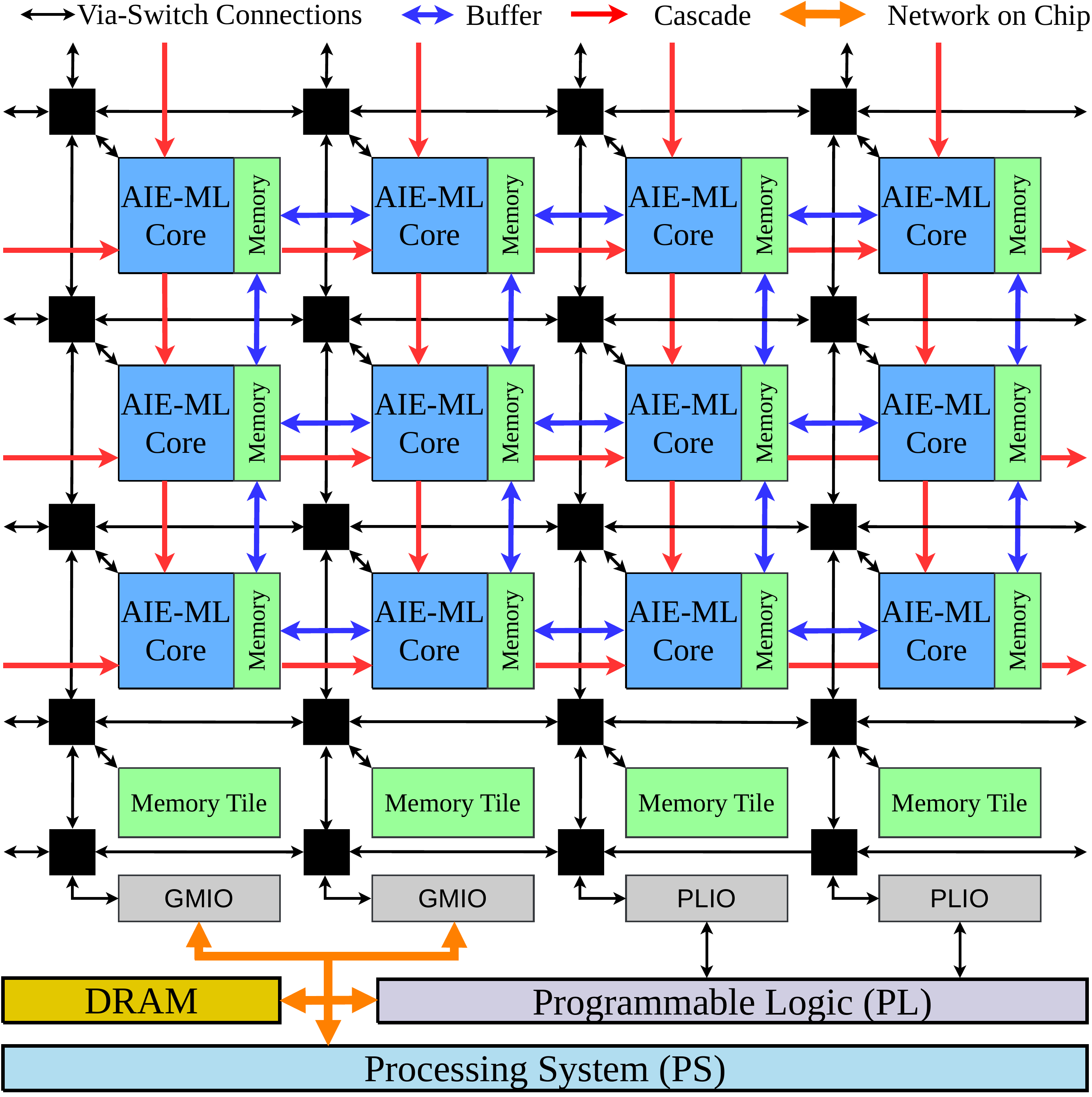}
  \caption{AMD Versal AIE-ML Architecture.}
  \label{fig:aie_ml_arch}
  \vspace{-5mm}
\end{figure}

The AMD Versal Adaptive Compute Acceleration Platform (ACAP) is a family of heterogeneous System-on-Chips (SoCs) designed to integrate different processing engines for high-performance computing (HPC) and AI workloads.
The architecture comprises the Processing System (PS), Programmable Logic (PL), and the AIE-ML  array.
\red{Versal also has a Network-on-Chip (NoC) that connects all these different components together to the DRAM.}
Figure \ref{fig:aie_ml_arch} shows an architecture diagram of Versal ACAP devices. 

\textbf{Processing System (PS) and Programmable Logic (PL):} The PS features a multi-core ARM processor responsible for host application execution, kernel scheduling, and managing the overall execution flow across the AIE and PL domains. 
The PL consists of traditional FPGA fabric, including Digital Signal Processors (DSPs), Block RAM (BRAMs), Look-Up Tables (LUTs), and Flip-Flops (FFs), providing hardware flexibility for custom logic and data pre-processing. 
    
\textbf{AIE-ML array:} 
The AIE-ML is a Very Long Instruction Word (VLIW) vector processor optimized for machine learning inference. 
Each AIE-ML tile contains a vector unit capable of significant throughput; for instance, it supports a vector width of 128 elements for bfloat16 (BF16), enabling up to 128 Multiply-Accumulate (MAC) operations per clock cycle. 
The architecture natively supports multiple precisions, including INT4, INT8, INT16, and INT32, with varying vector widths.
Each AIE-ML tile includes 64 KB of local data memory partitioned into four banks to facilitate concurrent access.
Every AIE-ML engine can run a separate kernel in parallel with all other AIEs.
In this work, we utilize the AMD Versal AI Edge VE2802 (on the VEK280 evaluation board), which features 304 AIE-ML engines arranged in a 38 $\times$ 8 grid.  
A critical feature of the AIE-ML architecture is the inclusion of \textbf{Memory Tiles}, with each tile providing 512 KB of localized, high-bandwidth storage. Each Memory Tile is organized into 16 single-port banks to minimize contention during complex data shuffling and tensor reshaping.
To facilitate rapid data movement, each memory tile is equipped with six input stream ports and six output stream ports. 
These ports allow the tile to simultaneously read from and write to the AIE-ML engines, ensuring that these engines are not stalled by memory bottlenecks.
Communication between the AIE array and other domains is facilitated by dedicated interface tiles.
There are 2 types of interface tiles: 
(1) Programmable Logic Input/Output (PLIO), which enables high-speed data transfer between the AIE-ML engines and the PL.
(2) Global Memory Input Output (GMIO), which provides a direct path between the AIE-ML engines and the external DRAM via the NoC. 

\rev{
Our implementation leverages the AIE-ML engines and Memory Tiles exclusively, avoiding any PL involvement. 
This design choice eliminates PL-to-AIE data transfers, keeping all computation and intermediate data within the AIE array for lower latency. 
It also leaves the PL available for complementary tasks such as real-time pre-processing (e.g., normalization and noise filtering of sensor data) and post-processing (e.g., driving output actuators), which are essential for edge deployment. 
Since all inference occurs within the AIE array, the external bandwidth requirements are limited to reading inputs and writing outputs. 
This low bandwidth demand allows us to use the GMIO interface for direct data movement between the AIE-ML engines and DRAM, without requiring PLIO.
}

\subsection{Versal AIE-ML Programmability}
The AIE-ML programming model consists of two primary abstractions: kernels and graphs. 
Kernels are the fundamental compute units, written in C/C++ using AMD's high-level APIs \cite{amd:2022:web:api_user_guide}.
Kernels are composed and interconnected through an Adaptive Data Flow (ADF) graph, which defines the structure and configuration of the AIE array. 
The ADF graph specifies the assignment of kernels to physical AIE engines, the input and output port connections for each kernel, and the configuration of off-chip interfaces. 
It also supports optional constraints such as kernel placement on specific engines, kernel co-location for time-sharing a single engine, and double buffering configuration. 
A subgraph represents a logical grouping of one or more kernels within the ADF graph that collectively implement a higher-level operation, such as a single network layer. 
This hierarchical organization enables modular design, where subgraphs can be independently developed, parameterized, and reused across different network architectures.

% Our implementation leverages the AIE-ML engines and Memory Tiles exclusively, utilizing the GMIO interface for direct data movement between the engines and DRAM. 
% While the PL offers additional acceleration potential, all operators in this study are deployed on the AIE-ML array to maximize performance for CRONet inference. 
% The PL can also be leveraged for low-computational-intensity tasks, such as real-time pre-processing and post-processing. 
% Pre-processing operations, such as normalization and noise filtering of sensor data, along with post-processing tasks, including driving output actuators, are essential when deploying this device at the edge. 

\section{Related work }

The application of topology optimization (TO) in structural health monitoring is frequently hindered by the huge computational overhead of iterative Finite Element Analysis 
% (FEA)
\cite{olabiyi:2025:cronet}. 
To mitigate these costs, recent research has explored hardware-specific acceleration and machine learning surrogates. 
Hesse et al. \cite{hesse:2023:to_fpga} introduced an application-specific instruction set processor (ASIP) implemented on an FPGA. 
While their approach utilized fixed-point arithmetic to streamline TO computations, 
% it remained unable to meet the stringent latency requirements of real-time applications.
\rev{their execution latency is very high (in tens of seconds).}
Neural architectures have emerged as a promising alternative to traditional numerical solvers. 
For instance, CRONet \cite{olabiyi:2025:cronet} employs a hybrid convolutional and recurrent neural network to approximate complex operator mappings, effectively serving as a high-speed surrogate for FEA. 
Although CRONet significantly reduces solver latency compared to traditional methods, \rev{it only addresses the problem at the algorithmic stage.
% achieving true real-time performance remains an challenging goal.
In contrast, our work operates at a different level of optimization, focusing on efficient hardware implementation rather than developing new models.
Thus, our contribution focuses on system-level acceleration of NN-based topology optimization to enable low-latency, energy-efficient execution.}

\begin{comment}
    
\red{TO, in the context of structural monitoring, comes under the umbrella of Digital Twin.}
The deployment of real-time digital twins on FPGA hardware has already demonstrated success in domains demanding high reliability and low latency, such as sodium-cooled nuclear reactors \cite{wang:2025:nuclear_dt} and power converters \cite{nwoke:2023:power_system_converter_dt}. 
These implementations not only eliminate cloud-based communication delays but also ensure data privacy at the edge. 
Furthermore, iterative ordinary differential equation (ODE) solvers are increasingly being replaced by hardware-optimized Neural ODEs and Gated Recurrent Units (GRUs) \cite{xu:2025:dt_edge_ai, xu:2025:fast_dt}. 
Building upon these advancements, this work addresses the remaining latency bottlenecks in TO by accelerating the CRONet architecture on AMD Versal AIE-ML engines, pushing the boundaries of real-time structural optimization.
\end{comment}

% \red{Fpga acceleration of CNN and RNN papers, Acceleration of CNN using AIE and then rest}

Since the debut of AMD’s AI Engine (AIE) architecture \cite{amd:web:versal}, a diverse range of workloads has been ported to the platform \cite{dong:2024:tcad:eq_vit, zhang:2022:fpl:hgcn, perryman:2023:ac:evaluation, singh:2023:ics:sparta, chen:2023:fpl:exploiting,yang:2023:iccad:aim,zhuang:2024:fpga:ssr,yemme:2023:ijcnn:scalable}. 
Central to these efforts is the optimization of GEMM, which serves as the computational backbone for most neural networks. 
Several frameworks have been proposed to map GEMM onto AIEs, each navigating the trade-offs between resource utilization and throughput.
CHARM \cite{zhuang:2023:fpga:charm, zhuang:2024:trets:charm_2} and AutoMM \cite{zhuang:2023:fpga:automm} prioritize resource efficiency for scalability, \rev{this results in high throughput but also results in high latency.}
In contrast, MaxEVA \cite{taka:2023:fpt:maxeva} maximizes performance at higher resource cost, separating multiplication and reduction into distinct kernels 
%—
allocating about 20\% of the AIE array solely for addition and capping efficiency at 80\%. 
Unlike MaxEVA’s shared memory transfer of partial sums, GAMA \cite{mhatre:2024:gama} unifies these operations within a single kernel, leveraging the high-bandwidth cascade interface to attain a 94\% active utilization of the AIE-ML array.
Other works, such as AMA \cite{deng:2024:fpl:ama}, attempt to boost performance through mixed-precision schemes (e.g., FP32-to-FP16). While effective for throughput, this approach introduces potential accuracy degradation.
More recently, RSN-XNN \cite{wang:2025:arxiv:rsn_xnn} introduced a reconfigurable stream network overlay to leverage the inherent heterogeneity of the Versal platform.

Wierse \cite{wierse:2023:ethz:evaluation} provides a detailed empirical analysis of the AI Engine’s communication interfaces, quantifying the bandwidth and latency trade-offs inherent in the interconnect. 
Complementing this, Mhatre et al. \cite{mhatre:2025:ispass:performance} investigate the sensitivity of GEMM performance to varying matrix dimensions and architectural parameters.

The challenge of AIE programmability has also spurred development in automated compilation toolchains. 
Vyasa \cite{chatarasi:2020:hpec:vyasa} extends the Halide DSL to generate AIE-compatible code, while ARIES \cite{zhuang:2025:fpga:aries} utilizes the MLIR framework \cite{lattner:2020:axriv:mlir} to target both Versal devices and Ryzen AI CPUs. 
Binder et al. \cite{elliott:2025:mm_aie_cpu} optimize AIE-ML matrix multiplication kernels by leveraging Memory Tiles to enhance data movement and locality on AMD's Ryzen AI CPUs.

While prior work on AMD Versal primarily focuses on accelerating individual kernels such as GEMM or specific neural network layers, these approaches do not address end-to-end deployment of complex DNNs with heterogeneous operators and strict data dependencies. In contrast, our work presents a full-system mapping of a hybrid CNN–RNN architecture for topology optimization, including custom operator implementations, dataflow-driven scheduling, and congestion-aware placement. This enables a complete on-AIE inference pipeline that minimizes off-chip communication and achieves low-latency, energy-efficient execution, which has not been demonstrated in prior Versal-based accelerators.

\section{Proposed architecture}
% \section{Proposed Methodology and Framework}
\begin{comment}
    
In this section, we first characterize the CRONet workload and highlight the parallelism opportunities exposed by its hybrid architecture. 
We then describe the constraints of the AMD Versal VEK280 platform and the resulting design challenges. 
Next, we present our bottom‑up design methodology: single AIE kernels for the major operators, the scaling strategy used to build multi AIE accelerators, and the layer fusing optimizations applied at both kernel and array levels. 
Finally, we describe the end‑to‑end dataflow and scheduling strategy that allows BranchNet and TrunkNet to execute in parallel on the AIE‑ML array.
\end{comment}

In this section, we first present an analysis of parallelism opportunities, data dependencies, and on-chip memory constraints for mapping CRONet on the AMD Versal AIE-ML architecture. We then present the overall dataflow execution model and scheduling strategy, followed by the proposed fusion techniques used to minimize data movement and improve efficiency. Next, we detail the implementation of key operators through custom kernels and scalable subgraph mappings, and introduce a reusable operator library for broader applicability. Finally, we present our custom placement strategy to address routing congestion and enable efficient deployment on the AIE-ML array.

\begin{figure*}[t]
  \centering
  \includegraphics[width=0.8\linewidth]{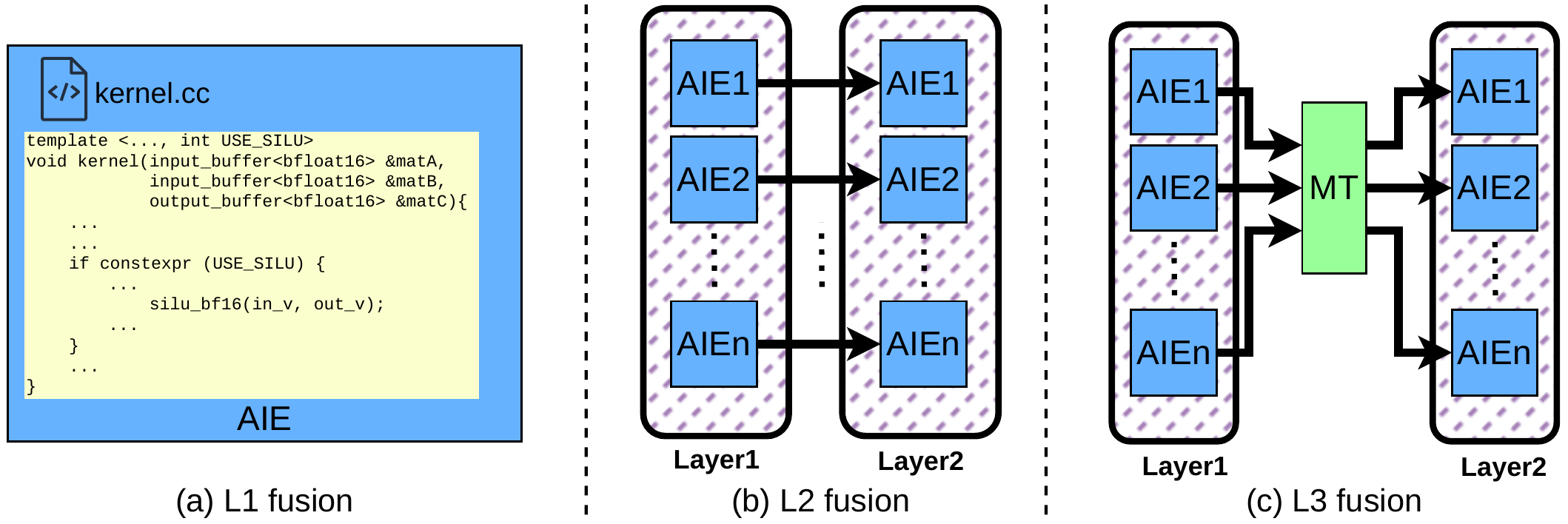}
  \caption{Different fusion techniques used in our implementation}
  \label{fig:fusion}
   \vspace{-6mm}
\end{figure*}

\subsection{Versal Architecture Mapping Analysis}

% Before implementation, it is essential to understand the structure of CRONet and the unique challenges it presents.
% CRONet is a hybrid neural network that combines both Convolutional Neural Network (CNN) and Recurrent Neural Network (RNN). 
% It comprises two sub-networks, \textbf{BranchNet} and \textbf{TrunkNet}, each designed to process distinct input streams. 
% The outputs of these two sub-networks are subsequently merged through an element-wise multiplication operation to generate a unified output.
\rev{As described in Section \ref{sec:background}, CRONet employs a dual-network architecture with distinct CNN and RNN components.
Mapping this architecture onto the AIE-ML array requires careful analysis of each layer's computational profile, memory footprint, and inter-layer data dependencies. 
% We begin by characterizing these requirements for both the BranchNet and TrunkNet, then identify the key mapping challenges that inform our design decisions.
}

\rev{
% A layer-by-layer breakdown of compute complexity (MACs) and memory requirements is summarized in Table \ref{tab:cronet_chr}. 
% For an input resolution of 20$\times$30, CRONet requires \textbf{53.5M MACs} and contains \textbf{419K} parameters (838 KB in BF16), with a total activation and weight footprint of 1.7 MB. 
% The original network operates in FP32, which we quantize to BF16 to better align with the Versal AIE-ML's native data types and maximize computational efficiency without loss of accuracy. 
The independent input streams of BranchNet and TrunkNet present a natural opportunity for parallel execution. 
However, conventional accelerators such as GPUs typically execute the two networks sequentially, leaving significant compute resources idle during each sub-network's execution.}

% A layer‑by‑layer breakdown with compute complexity (MACs) and memory requirements is summarized in Table \ref{tab:cronet_chr}. 
% Overall, CRONet requires \textbf{105.8M MACs} and contains \textbf{419 K} parameters (838 KB in BF16) and totoal activation + weights footprint is around 2.5MB for a size of 20x60. \rev{Need to reword this..}
\rev{Furthermore, the iterative nature of topology optimization introduces a strict sequential dependency where the complete output of one iteration is processed and fed as input to the next. 
This makes batch processing infeasible, severely limiting GPU throughput, as GPUs rely heavily on large batch sizes to amortize overhead and saturate their compute units.}
% This hybrid architecture presents a valuable opportunity for parallel execution and performance optimization. However, conventional hardware accelerators such as GPUs, typically execute the two networks sequentially, leading to resource underutilization and suboptimal performance.
In contrast, the AIE-ML array's dataflow-oriented architecture provides fine-grained control over individual engines, allowing them to be interconnected in custom topologies tailored to the network structure. 
This enables us to execute BranchNet and TrunkNet in parallel, directly addressing the GPU's sequential execution bottleneck and achieving low-latency inference.
\rev{However, this comes at the cost of significant design complexity. 
Mapping computation across the AIE array in a way that is both correct and efficient requires careful planning of data movement and synchronization through the interconnect.}

% \begin{table}[t]\centering
% \setlength{\tabcolsep}{9pt}
% \renewcommand{\arraystretch}{1}
% \caption{CRONet characterization showing compute and memory requirements of the network. A=Activation , W=Weights in BF16 for 30x20}
% \label{tab:cronet_chr}
% %\resizebox{\textwidth}{!}{ % use this if the table is too large
% \begin{tabular}{lccccc}\toprule
% Network &Layer &Parameters & \makecell{Compute \\ (MACs)} &\makecell{Memory \\ (A+W)} \\\midrule
% \multirow{4}{*}{TrunkNet} &CONV3D &288 &562K &63.1K \\
% &CONV3D &9K &24M &285K \\
% &Linear &192K &192K &384K \\
% &Linear &102K &102K &210K \\
% \hline \\
% \multirow{5}{*}{BranchNet} &CONV2D &144 &864K &192K \\
% &CONV2D &4.6K &27.6M &393K \\
% &RNN & 6.1K & 61.4K & 14.8K \\
% &Linear &2.5K &2.5K &5.2K \\
% &Linear &102K &102K &210K \\
% \hline \\
%  &Total & 419K & 53.5M & 1.7M \\
% \bottomrule
% \end{tabular}
% \end{table}

% The AIE-ML array consists of a two-dimensional grid of AIE-ML Engines. 
% It uses the GMIO and PLIO interfaces to communicate between PL and DRAM.
% In this paper, we utilize the GMIO interface exclusively for transferring data between AIE-ML and DRAM.

\rev{The limited DRAM bandwidth available on the Versal platform makes off-chip data movement during inference particularly costly.
This reinforces our design decision to retain all intermediate activations on-chip, either in AIE local memory or in Memory Tiles, restricting DRAM access to input reads and output writes only.
All compute layer weights are stored as constants within AIE local memory, as their relatively small total size makes on-chip storage feasible, eliminating repeated weight fetching from external memory and further reducing communication overhead during inference.
}

\subsection{Overall CRONet network dataflow and scheduling}
\rev{The overall CRONet execution is organized as a dataflow pipeline that maximizes parallelism across the AIE-ML array. 
BranchNet and TrunkNet, along with all their constituent layers, are fully implemented on the AIE array with no reliance on the PL. 
All network weights are stored permanently in AIE local memory, eliminating weight transfers from DRAM during inference. 
Input data for both sub-networks is streamed from DRAM into the AIE array via the GMIO interface, and the final output is streamed back upon completion. 
All intermediate activations remain on-chip throughout execution through L1, L2, or L3 fusion, ensuring that no data is written back to DRAM at any point during inference.
% The current implementation utilizes up to 73\% of the available AIE resources.
% As a future work, a hybrid execution approach can be explored where element-wise operations such as SiLU are offloaded to the PL fabric, freeing AIE resources for more compute-intensive operations and potentially improving overall resource utilization and throughput.
}

\subsection{Fusion strategies}
\rev{
To enable efficient dataflow and minimize off-chip data movement, we develop three fusion strategies, illustrated in Figure \ref{fig:fusion}. 
\textbf{L1 fusion} integrates element-wise operations directly into compute kernels, eliminating intermediate memory traffic between successive operations. 
\textbf{L2 fusion} streams outputs between adjacent subgraphs through direct AIE-to-AIE buffer connections, enabling tightly pipelined execution without routing data through Memory Tiles. 
\textbf{L3 fusion} routes intermediate data through Memory Tiles either to buffer activations that exceed AIE local memory capacity or to perform required data reshaping before the next stage of computation.
}

\subsubsection{L1 fusion}
\rev{L1 fusion operates at the operator level, where element-wise operations are fused directly with compute kernels to form a single composite kernel. 
For example, the SiLU activation function is fused into all convolution and GEMM kernels, eliminating the need to write intermediate results back to memory and re-read them in a separate kernel. 
This reduces both AIE resource consumption and memory traffic. The fusion is configurable through a template parameter, allowing it to be enabled or disabled across different layer types.
}

\subsubsection{L2 Fusion}
% subgraph 
\rev{L2 fusion operates at the layer level, where the output of one subgraph is streamed directly to the next through AIE-to-AIE buffer connections, bypassing Memory Tiles and DRAM entirely. 
This direct inter-subgraph dataflow reduces latency and enables a tightly coupled execution pipeline across the array. 
While L1 and L2 fusion are well-established techniques in deep learning compilers, to the best of our knowledge, their implementation within the AIE-ML programming model has not been previously explored, making this a novel contribution of our work.
}

\subsubsection{L3 Fusion}
\rev{L3 fusion operates at a higher level, handling cases where intermediate data exceeds AIE local memory capacity or requires reshaping before the next stage. 
In such cases, the output of a subgraph is routed through a Memory Tile, which serves as a staging buffer between subgraphs. 
While this introduces an additional hop compared to L2 fusion, it avoids costly DRAM accesses and keeps the data within the on-chip memory hierarchy.}

\begin{figure*}[t]
  \centering
  \includegraphics[width=1
  \linewidth]{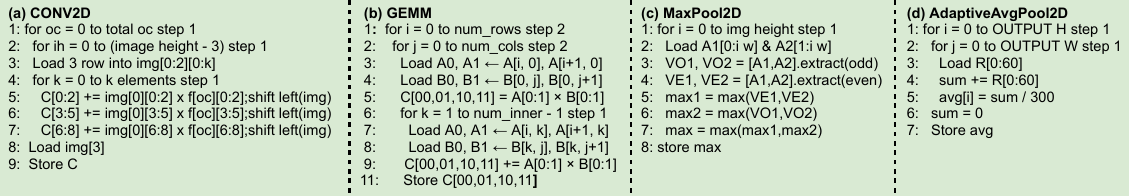}
  \caption{Pseudocode for AIE-ML kernel implementation for various operators}
  \label{fig:six_algorithms}
   \vspace{-6mm}
  %\vspace{-4mm}
\end{figure*}
% \subsection{Single AIE kernel implementation}
\subsection{Operator Implementation}
\begin{figure*}[t]
  \centering
  \includegraphics[width=1
  \linewidth]{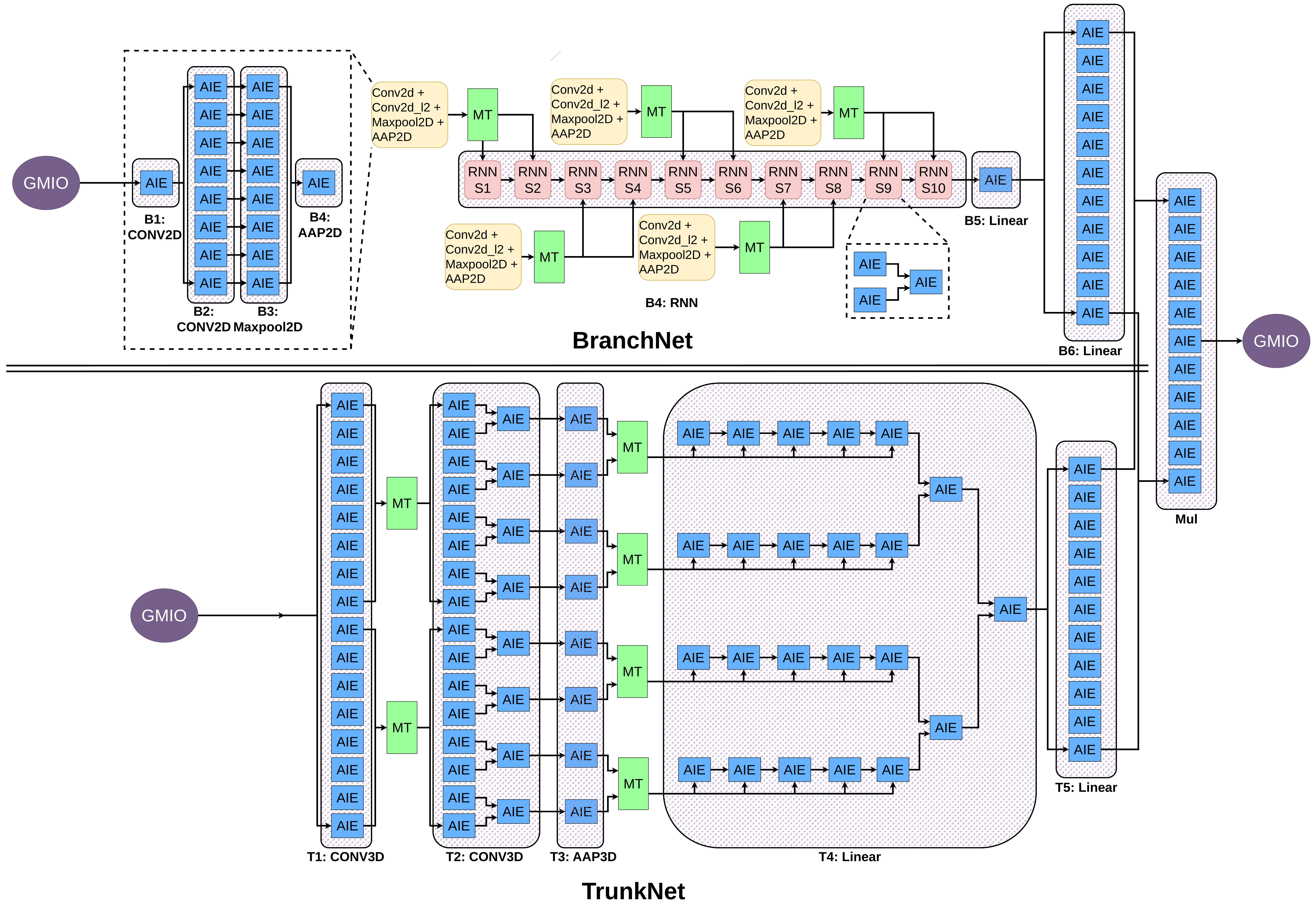}
  \caption{The ADF graph of our CRONet implementation}
  \label{fig:cronet_adf_graph}
  \vspace{-4mm}
\end{figure*}

In this section, we present efficient kernel implementations and their corresponding subgraph mappings for each operator in CRONet. While prior work on Versal has addressed standard operations such as GEMM and activation functions, CRONet introduces operators that have not been previously mapped to the AIE-ML architecture, including 3D convolution and adaptive average pooling. These operators pose distinct implementation challenges due to their dynamic memory access patterns and unique processing requirements. We develop custom, highly optimized kernels for each, as illustrated in Figure \ref{fig:six_algorithms}, which presents pseudocode for the convolution, GEMM, max pooling, and adaptive average pooling algorithms. SiLU implementation is adopted from prior work \cite{AMD:2025:git:iron_tollkit}. Each kernel is further encapsulated within a modular, parameterizable subgraph that defines the spatial connectivity and dataflow across multiple AIE-ML engines, enabling layer-level scaling across the available resources.
Figure \ref{fig:cronet_adf_graph} shows the complete mapping of the TrunkNet and BranchNet network. 
% In this section, we present efficient kernel implementations for each operator in CRONet. 
% While prior work on Versal has addressed standard operations such as GEMM and activation functions, CRONet introduces operators that have not been previously mapped to the AIE-ML architecture, including 3D convolution and adaptive average pooling. 
% These operators pose distinct implementation challenges due to their dynamic memory access patterns and unique processing requirements. 
% We develop custom, highly optimized kernels for each, as illustrated in Figure \ref{fig:six_algorithms}, which presents pseudocode for the convolution, GEMM, max pooling, and adaptive average pooling algorithms. SiLU implementation is adopted from prior work \cite{AMD:2025:git:iron_tollkit}. 

\subsubsection{Convolution 2D/3D}
The convolution kernel efficiently maps the convolution operation onto the AIE-ML vector processor using a vectorized sliding window approach. Each filter element is multiplied across the input width in a single vector operation. The input is then shifted by one position and multiplied with the next filter element, accumulating partial results. This process repeats across all filter dimensions to produce the final convolution output. Computation is parallelized along the width dimension of the input feature map, enabling multiple output elements to be computed simultaneously and maximizing utilization of the AIE vector processing units. The kernel naturally extends to support 3D convolution by introducing an additional depth dimension to the sliding window computation.

The kernel is fully parameterized through template parameters, including input dimensions, filter size, and the number of input and output channels, enabling a single implementation to support a wide range of layer configurations. All parameters are bounded by the available AIE local memory, requiring a trade-off between spatial resolution and channel parallelism.

Conv3D in TrunkNet and Conv2D in BranchNet are each mapped as independent subgraphs. When a layer's memory requirements exceed AIE local memory capacity due to larger input dimensions or higher channel counts, the workload is distributed across multiple AIEs. The convolution subgraph supports partitioning along input channels, output channels, or both, with all options exposed as parameters selectable at instantiation time without modifying the underlying subgraph design.

\subsubsection{General Matrix Multiplication (GEMM)}
The GEMM kernel builds upon GAMA \cite{mhatre:2024:gama}, a state-of-the-art framework for GEMM implementation on the Versal AIE-ML. While GAMA provides efficient matrix multiplication, it operates with fixed kernel dimensions and streams both input matrices from external memory. We extend GAMA in two ways. First, the weight matrix is stored persistently in AIE local memory rather than streamed from DRAM, eliminating excess DRAM reads. Second, we introduce full parameterization of the M, K, and N matrix dimensions, enabling support for the highly asymmetric GEMM sizes present across different layers of CRONet. Internally, the kernel follows a tiled inner-product strategy, decomposing the multiplication into smaller operations over a base tile size of 8$\times$8$\times$4 using AMD's MMUL API \cite{amd:2022:web:api_user_guide}.

Rather than relying on PLIO-based data movement as in GAMA, input data is sourced either directly from Memory Tiles or from AIE local memory, enabling tighter integration within the on-chip dataflow pipeline. Each GEMM layer across TrunkNet and BranchNet is mapped to a dedicated, independently parameterized subgraph, allowing custom configuration per layer.
A key limitation of GAMA's scaling strategy is that the K dimension is constrained by the device's cascade chain limit, which is 38 columns on the VEK280. Mapping a complete subgraph across all available columns introduces severe routing congestion, and for sufficiently large K values, the workload may not fit within this limit at all. To overcome this, large K dimensions are sliced across multiple independent AIE clusters, each utilizing cascade chaining internally, with partial results reduced via an adder tree. This decouples scalability from the physical cascade chain limit and alleviates placement congestion, enabling support for highly asymmetric GEMM sizes that would otherwise be infeasible on the device.

\subsubsection{RNN}
\rev{The RNN layer is implemented by reusing the GEMM kernels without requiring a dedicated RNN kernel. 
The RNN is fully unrolled within its subgraph, with each time step mapped as an independent computation on the AIE array, enabling the entire RNN execution to be expressed as a static dataflow graph. 
Each time step additionally includes an add kernel, with the Tanh activation function fused directly into it via L1 fusion.
% All RNN weights are stored persistently in AIE local memory, eliminating repeated weight transfers from DRAM across the sequential steps of the unrolled computation.
}

\subsubsection{Other layers}

\textbf{MaxPool2D:}
\rev{The kernel extracts elements within each pooling region into vectors using AIE APIs such as \textit{aie::filter\_even}, \textit{aie::filter\_odd}, and \textit{aie::shuffle}, then computes the maximum across the resulting vectors, yielding a fully vectorized pooling operation. 
The kernel is fully parameterized, exposing input height, width, and channel count as template parameters to support varying layer configurations.
}

\textbf{AdaptiveAveragePool 2D/3D:}
\rev{The Adaptive Average Pooling (AAP) kernel dynamically adjusts its pooling regions and strides to produce a fixed-size output regardless of input dimensions. 
Unlike fixed pooling operators, AAP must handle variable stride lengths and pooling windows for a given input, resulting in irregular memory access patterns that complicate vectorization. 
The kernel processes the input feature map row by row, extracting pooling regions into vectors and advancing sequentially across rows in a sliding window fashion. 
The extracted vectors are then averaged using vectorized operations. The AAP3D variant extends this approach by operating across an additional depth dimension. 
Input and output spatial dimensions, along with channel count, are exposed as template parameters for layer adaptability.}

\textbf{Sigmoid Linear Unit:}
The Sigmoid Linear Unit (SiLU) activation function is implemented as a lightweight, low-latency kernel designed for efficient nonlinear activation. 
Direct computation of the sigmoid function is computationally expensive for vector processors. Therefore, this kernel employs a lookup table (LUT) based implementation, adapted from the IRON toolkit SiLU implementation \cite{AMD:2025:git:iron_tollkit}.

% Figure \ref{fig:six_algorithms} (f) has a pseudocode for SiLU.

% \subsection{Mapping and scaling strategy}
% \rev{Building upon the individual kernel implementations described in the previous section, we construct modular, parameterizable subgraphs that define the spatial connectivity and dataflow across multiple AIE-ML engines. 
% Each network layer in CRONet is mapped as an independent subgraph onto the AIE-ML array, enabling layer-level parameterization and flexible scaling across the available resources.}

\subsection{Reusable operator library}
\rev{The parameterized kernels and their corresponding subgraphs together form a reusable operator library for the Versal AIE-ML array. 
Each operator is independently configurable through template parameters, allowing developers to instantiate and integrate individual layers into new network designs with minimal modification. 
The library is designed to be network-agnostic, providing foundational building blocks for deploying a broad range of neural network architectures on the Versal AIE-ML platform beyond CRONet. 
In its current form, the library, which will be open-sourced, covers all operators required by CRONet, with expanded operator coverage planned as future work.
}

\subsection{Custom placement}
\rev{
To achieve high performance, the CRONet mapping uses a resource-intensive implementation, particularly in terms of the number of AIEs used. At such high utilization, the default compiler placement fails
to produce a valid routing solution, even at a high compiler optimization level.
Thus, we design a simple custom placement algorithm. 
Kernels that communicate frequently are placed in close physical proximity on the AIE array, minimizing wire lengths and reducing routing pressure at congestion-prone junctions. 
Specifically, the algorithm analyzes the connectivity between subgraphs and assigns kernel locations such that dataflow-adjacent kernels occupy neighboring engines, aligning physical placement with logical dataflow locality. 
This eliminates the long cross-array routes that cause congestion under automatic placement.
}
\section{Methodology}
\begin{table*}[t]\centering
\setlength{\tabcolsep}{6pt}
\renewcommand{\arraystretch}{1}
\caption{CRONet characterization showing compute and memory requirements of the network. A=Activation, W=Weights in BF16}
\label{tab:cronet_chr}
\begin{tabular}{lcccccccc}\toprule
\multicolumn{3}{c}{\textbf{CRONet Size}} &\multicolumn{2}{c}{\textbf{30$\times$10}} &\multicolumn{2}{c}{\textbf{30$\times$20}} &\multicolumn{2}{c}{\textbf{60$\times$20}} \\
\hline
Network &Layer &Parameters &\makecell{Compute \\ (MACs)} &\makecell{Memory \\ (A+W)} &\makecell{Compute \\ (MACs)} &\makecell{Memory \\ (A+W)} &\makecell{Compute \\ (MACs)} &\makecell{Memory \\ (A+W)} \\\midrule
\multirow{4}{*}{TrunkNet} &CONV3D &288 &294K &33.3KB &562K &63.1KB &1.1M &123KB \\
&CONV3D &9K &12.6M &105KB &24M &285KB &47.2M &346KB \\
&Linear &192K &192K &384KB &192K &384KB &192K &384KB \\
&Linear &102K &102K &210KB &102K &210KB &102K &210KB \\
\hline \\
\multirow{5}{*}{BranchNet} &CONV2D &144 &432K &96KB &864K &192KB &1.7M &384KB \\
&CONV2D &4.6K &13.8M &201KB &27.6M &393KB &55.3M &777KB \\
&RNN &6.1K &61.4K &14.8KB &61.4K &14.8KB &61.4K &14.8KB \\
&Linear &2.5K &2.5K &5.2KB &2.5K &5.2KB &2.5K &5.2KB \\
&Linear &102K &102K &210KB &102K &210KB &102K &210KB \\
\hline \\
 &Total &419K &27.6M &1.3MB &53.5M &1.7MB &105.8M &2.5MB \\
\bottomrule
\end{tabular}
\end{table*}
\noindent{\textbf{Target Platform:}}
All designs are evaluated on the AMD Versal VEK280 Evaluation Kit, compiled and implemented using AMD Vitis 2025.2. 
Performance results are obtained by executing the designs directly on the VEK280 board using AMD's default platform. 
Power measurements are performed using the \texttt{system controller} application \cite{amd:2023:web:sc}, which interfaces with the board's built-in PMIC. 
Rail-level power readings are collected for all active rails and summed to obtain the total power consumption of the device during inference.

\noindent{\textbf{Baseline:}}
\rev{Versal performance is evaluated against an Nvidia T4 GPU, a device designed for inference workloads within a comparable power envelope with a TDP of 75W. 
Table \ref{tab:devices} lists the devices and their respective Tech node and TDP.
To ensure a fair comparison, T4 performance is normalized to the 7nm process node used by the Versal platform, following the scaling methodology of Stillmaker et al \cite{stillmaker:2017:scaling}. 
PyTorch-based execution is used for GPU baselines. 
GPU power is measured using nvidia-smi.}

\noindent{\textbf{Precision:}}
\rev{CRONet is trained in FP32 precision. 
For inference on Versal, reduced precision is explored to improve performance. 
BF16 is selected as the inference precision, as it does not degrade solution quality, whereas INT8 was found to reduce accuracy. 
For GPU execution, BF16 is not natively supported on the T4, so FP16 was evaluated as an alternative. 
However, FP16 yields poor performance at a batch size of 1 compared to FP32, while higher batch sizes show improved FP16 performance. 
We compare BF16 on Versal to FP32 on T4 due to a lack of efficient FP16/BF16 support at batch size 1. 
This reflects practical deployment constraints rather than peak GPU capability.
A batch size of 1 is used consistently across all experiments.}

\noindent{\textbf{Workload:}}
The baseline code for CRONet \cite{olabiyi:2025:cronet} was run with different material distributions. We run three different sizes.
We double the amount of elements in the material distribution by 2x every time starting from a distribution of 30x10 (small), 30x20 (medium) and 60x20 (large). 
Unless specified, all results are reported with the medium material distribution.

\begin{table}[t]
    \centering
    \caption{Specifications of hardware platforms used for evaluation.}
    \label{tab:devices}
    \begin{tabular}{>{\raggedright\arraybackslash}lccc}
        \toprule
        \textbf{Devices} 
        &\makecell{\textbf{GPU}} 
        &\makecell{\textbf{Versal}} \\\midrule
        Name & Nvidia T4 & AMD VEK280 \\
        Tech Node (nm) &14 &7 \\
        Thermal Design Power (TDP) & 75 W & 75 W \\
        \bottomrule
    \end{tabular}
\end{table}

% \begin{figure}[t]
%   \centering
%   \includegraphics[width=1
%   \linewidth]{fig/CRONet_array.png}
%   \caption{CRONet subgraphs of TrunkNet (T1 to T5) and BranchNet (B1 to B7) mapped to the physical location on the AIE-ML array}
%   \label{fig:array_map}
%   % \vspace{-3mm}
% \end{figure}

% \noindent{\textbf{CRONet Setup}:}
% CRONet is neural network model based on DeepONet \cite{lu:2021:deeponet} family of models.
% \red{
% % Our implementation utilizes a batch size of 3, with the original 
% We train CRONet using FP32 precision. 
% For inference on the Versal platform, we employ BF16 precision. 
% BF16 precision has been verified within the topology optimization application flow, demonstrating similar convergence behavior without any degradation in solution quality.
% Conversely, we maintain FP32 precision for both CPU and GPU execution, as neither supports the BF16 format natively.
% The CRONet application flow requires a batch size of 1 for all inference tasks; therefore, all results are compared using a batch size of 1.}

\noindent{\textbf{Performance Metrics:}}
\rev{Two metrics are used to evaluate the performance of the deployed solution. 
\textbf{Inference latency}, measured in milliseconds, is used to assess the execution time of the CRONet model across all platforms. 
\textbf{Energy efficiency}, quantified as the number of inferences per Watt, is used to evaluate the power-performance trade-off and provide a complete picture of the suitability of each platform for deployment in realistic environments.}

\begin{table}[t]
    \centering
    \caption{Impact of inference precision on CRONet solution accuracy in a 100-iteration hybrid NN-FEA topology optimization experiment.}
    \label{tab:precision}
    \begin{tabular}{lcc}
        \toprule
        \textbf{Precision} & \makecell{\textbf{CRONet Invocations} \\ \textbf{(out of 100)}} & \makecell{\textbf{Solution} \\ \textbf{Accuracy}} \\
        \midrule
        FP32  & 33 & 100.00\% \\
        BF16  & 33 & 100.00\% \\
        INT8  & 30 &  90.91\% \\
        \bottomrule
    \end{tabular}
\end{table}

\section{Results}

% \begin{table*}[t]
% \centering
% \small
% \caption{Latency, power, and energy efficiency comparison of CRONet across material distribution sizes (30$\times$10, 30$\times$20, 60$\times$20) on Versal VEK280 and Nvidia T4.}
% \label{tab:main_perf}
% \begin{tabular}{c|ccc|ccc|cc}
% \hline
% \textbf{CRONet Size} &\multicolumn{3}{c|}{\textbf{VEK280}} &\multicolumn{3}{c|}{\textbf{T4}} &\multicolumn{2}{c}{\textbf{Improvement over GPU}} \\
% \hline
% \textbf{X * Y} &\textbf{Latency (ms)} &\textbf{Power (W)} &\makecell{\textbf{Energy} \\ \textbf{Efficiency}} &\textbf{Latency (ms)*} &\textbf{Power (W)} &\makecell{\textbf{Energy} \\ \textbf{Efficiency}} &\textbf{Latency} &\makecell{\textbf{Energy} \\ \textbf{Efficiency}} \\
% \hline
% 30x10 &0.45 &19.06 &116.59 &1.12 &32 &27.90 &2.49 &4.18 \\
% 30x20 &0.52 &21.15 &90.93 &1.19 &35 &24.01 &2.29 &3.79 \\
% 60x20 &0.82 &21.49 &56.75 &1.25 &37 &21.62 &1.52 &2.62 \\
% \hline
% \multicolumn{9}{l}{\footnotesize *T4 latency is normalized to 7nm.} \\
% \end{tabular}
% \end{table*}

\begin{table}[t]
\centering
\small
\caption{AIE-ML engine, Memory Tile, and GMIO channel allocation per layer for CRONet on the Versal VEK280.}
\label{tab:versal_resource_util}
\begin{tabular}{lcccc}
\toprule
\textbf{Network} &\textbf{Layer} &\textbf{AIEs} &\textbf{Memtile} &\textbf{GMIO} \\\midrule
\multirow{5}{*}{TrunkNet} &CONV3D &16 &\multirow{5}{*}{6} &\multirow{5}{*}{1} \\
&CONV3D &24 & & \\
&AAP3D &8 & & \\
&Linear &23 & & \\
&Linear &11 & & \\
\hline
\multirow{7}{*}{BranchNet} &CONV2D &5 &\multirow{7}{*}{5} &\multirow{7}{*}{5} \\
&CONV2D &40 & & \\
&MaxPool2D &40 & & \\
&AAP2D &5 & & \\
&RNN &28 & & \\
&Linear &1 & & \\
&Linear &11 & & \\
\hline
Mul & &11 & &11 \\
\hline
&Total &223 &11 &17 \\
\hline
&Total \% &73\% &14\% &35\% \\
\bottomrule
\end{tabular}
\end{table}

\subsection{\rev{Impact of Quantization on Solution Accuracy}}
To identify the optimal inference precision for AIE-ML deployment, we evaluate CRONet under three numerical precisions: FP32, BF16, and INT8. 
Accuracy is assessed using a hybrid NN-FEA execution strategy over 100 optimization iterations, where each iteration dynamically selects between FEA and CRONet based on the error of the previous iteration's output \cite{olabiyi:2025:cronet}. 
If the error exceeds a defined threshold, FEA is invoked; otherwise, CRONet is used as a computationally efficient surrogate. 
Table \ref{tab:precision} summarizes the results.
Both FP32 and BF16 achieve 100\% solution accuracy, with CRONet selected for 33 of the 100 iterations in each case, confirming that BF16 quantization introduces no degradation in solution quality. 
In contrast, INT8 reduces accuracy to 90.91\%, with FEA invoked more frequently (70 out of 100 iterations), indicating that the aggressive quantization compromises the network's predictive fidelity. 
Based on these results, we select BF16 as the inference precision, as it aligns with the AIE-ML's native data types and delivers full accuracy while enabling higher computational throughput than FP32.

\subsection{\rev{Workload Characterization Results}}
A layer-by-layer breakdown of compute complexity (MACs) and memory requirements across all three material distribution sizes is summarized in Table \ref{tab:cronet_chr}. 
For the smallest configuration (30$\times$10), CRONet requires 27.6M MACs and contains 419K parameters (838 KB in BF16), with a total activation and weight footprint of 1.3 MB. As the material distribution size increases to 30$\times$20 and 60$\times$20, the compute and memory requirements scale accordingly, with MACs increasing to 53.5M and 105.8M and the total activation and weight footprint growing to 1.7 MB and 2.5 MB, respectively. 
The parameter count remains constant at 419K across all sizes, as only the activation dimensions change with input size. 

\subsection{\rev{End-to-End Performance Comparison}}
\begin{figure}[t]
  \centering
  \includegraphics[width=1
  \linewidth]{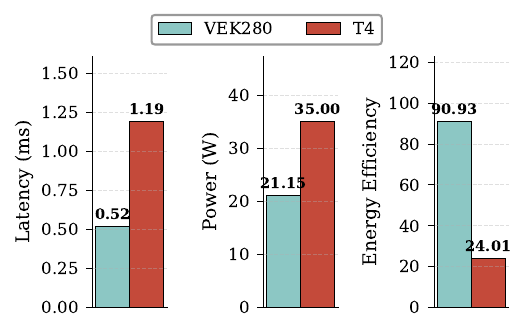}
  \caption{Performance comparison of CRONet (30$\times$20) inference on Versal VEK280 and Nvidia T4 across latency, power consumption, and energy efficiency.}
  \label{fig:perf_compare_30x20}
  \vspace{-4mm}
  % \vspace{-3mm}
\end{figure}

\begin{figure}[t]
  \centering
  \includegraphics[width=1
  \linewidth]{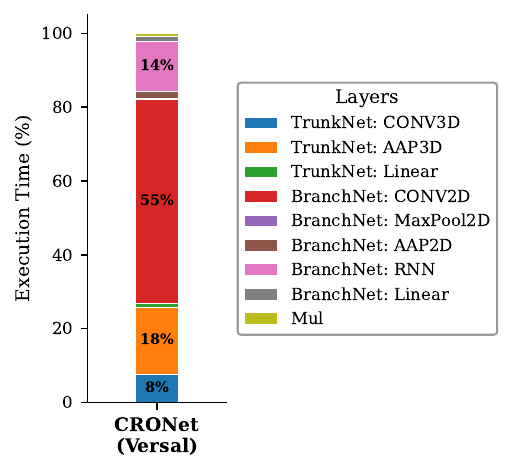}
  \caption{Layer-wise percentage breakdown of CRONet inference execution time on the Versal VEK280.}
  \label{fig:cronet_exe_breakdown}
  \vspace{-4mm}
  % \vspace{-3mm}
\end{figure}

Figure \ref{fig:perf_compare_30x20} provides a detailed comparison of latency, power consumption, and energy efficiency between the Versal VEK280 and the Nvidia T4, an inference-class ML-optimized GPU.
The CRONet implementation on Versal achieves a latency of 0.52 ms compared to 1.19 ms on the T4, a 2.29$\times$ improvement. 
This is attributed to our custom dataflow implementation, which enables parallel execution of BranchNet and TrunkNet and leverages L1, L2, and L3 fusion to eliminate DRAM access for intermediate data.
Additionally, TO’s iterative algorithm must be executed with a batch size of 1, which prevents the GPU from reaching peak performance, as GPUs depend on large batch sizes to fully utilize their computational resources.
Power consumption is also notably lower at 21.15 W versus 35.00 W for the T4, reflecting the inherent power efficiency of the AIE-ML architecture, which further benefits from reduced off-chip data movement. 
Together, these translate to a 3.79$\times$ improvement in energy efficiency, as the lower latency and reduced power compound to deliver significantly more energy efficiency. 

\subsection{\rev{Performance Breakdown}}
To identify performance bottlenecks across the network and guide future optimization efforts, we analyze the layer-wise execution time breakdown of CRONet on Versal, presented in Figure \ref{fig:cronet_exe_breakdown}. 
The CONV2D layer in BranchNet dominates execution time at 55.3\%, where irregular memory access patterns prevent full utilization of the VLIW instruction width on the AIE-ML vector processor, compounded by its high computational demand as reflected in the MAC count in Table \ref{tab:cronet_chr}. 
In contrast, the CONV3D layer has similar computational requirements but lower memory demands, making it significantly faster than CONV2D.
The AAP3D layer in TrunkNet is the next largest contributor at 18.1\%. 
The RNN layer, despite its sequential nature, executes efficiently through L1 fusion, where the output of each sequence step is directly fed to the next without returning to external memory. 
Similarly, linear and pooling layers exhibit low execution times, reflecting their modest computational and memory requirements. 
The remaining layers, including the final element-wise multiplication, collectively account for a negligible fraction of the total execution time. 
CONV2D and AAP3D remain the primary bottlenecks, and further optimization through improved memory access patterns and enhanced vectorization is left as future work.

\begin{table*}[t]
\centering
\small
\caption{Latency, power, and energy efficiency comparison of CRONet across material distribution sizes (30$\times$10, 30$\times$20, 60$\times$20) on Versal VEK280 and Nvidia T4.}
\label{tab:main_perf}
\begin{tabular}{c|ccc|ccc|cc}
\hline
\textbf{CRONet Size} &\multicolumn{3}{c|}{\textbf{VEK280}} &\multicolumn{3}{c|}{\textbf{T4}} &\multicolumn{2}{c}{\textbf{Improvement over GPU}} \\
\hline
\textbf{X * Y} &\textbf{Latency (ms)} &\textbf{Power (W)} &\makecell{\textbf{Energy} \\ \textbf{Efficiency}} &\textbf{Latency (ms)*} &\textbf{Power (W)} &\makecell{\textbf{Energy} \\ \textbf{Efficiency}} &\textbf{Latency} &\makecell{\textbf{Energy} \\ \textbf{Efficiency}} \\
\hline
30x10 &0.45 &19.06 &116.59 &1.12 &32 &27.90 &2.49 &4.18 \\
30x20 &0.52 &21.15 &90.93 &1.19 &35 &24.01 &2.29 &3.79 \\
60x20 &0.82 &21.49 &56.75 &1.25 &37 &21.62 &1.52 &2.62 \\
\hline
\multicolumn{9}{l}{\footnotesize *T4 latency is normalized to 7nm.} \\
\end{tabular}
\vspace{-4mm}
\end{table*}

\subsection{\rev{Resource Utilization}}
Table \ref{tab:versal_resource_util} details the resource utilization on the AMD Versal VEK280 platform. Our implementation utilizes 223 of the 304 available AIE-ML engines (73\%), 11 Memory Tiles, and 17 GMIO channels. 
The majority of AIE engines are allocated to convolution layers, with the two CONV2D layers in BranchNet alone consuming 45 engines and the two CONV3D layers in TrunkNet requiring 40. 
MaxPool2D is the next most resource-intensive layer at 40 engines, driven by the large output volume of the preceding CONV2D layers.  
Linear and RNN layers require a comparatively smaller footprint. Notably, BranchNet consumes a larger share of the total AIE resources (134 engines) compared to TrunkNet (78 engines), reflecting its greater compute complexity in a hybrid CNN-RNN network.
Despite this imbalance, both sub-networks execute in parallel, with the faster TrunkNet completing execution while BranchNet continues processing. 
Memory Tiles serve as staging buffers for intermediate activations between layers, while GMIO channels are distributed across both sub-networks and the final element-wise multiplication node to facilitate data movement between the AIE array and DRAM.

% Table \ref{tab:versal_resource_util} details the resource consumption on the AMD Versal VEK280 platform. 
% Our implementation utilizes 73\% of the available AIE-ML engines, with the majority allocated to convolution operations, while non-linear operations (such as pooling and activations) require a significantly smaller AIE footprint.

\subsection{\rev{Scaling}}
To evaluate the scalability of our implementation, we assess CRONet performance across three input material distributions: 30$\times$10, 30$\times$20, and 60$\times$20, where each configuration represents a 2x increase in problem size. 
Table \ref{tab:main_perf} summarizes the latency, power consumption, and energy efficiency for both the Versal VEK280 and the Nvidia T4 across all three configurations.
On Versal, latency scales from 0.45 ms at 30$\times$10 to 0.82 ms at 60$\times$20, an increase of only 1.82$\times$ despite a 4$\times$ growth in input size. 
This sub-linear scaling demonstrates the effectiveness of our dataflow implementation in absorbing additional computation through parallel execution and on-chip memory fusion. 
Power consumption remains nearly constant across all three configurations (19.06–21.49 W), indicating that the AIE-ML array's power draw is largely independent of input size within this range.
In contrast, the T4 exhibits relatively flat latency scaling (1.12–1.25 ms) due to consistent underutilization at batch size 1, but at a significantly higher power cost (32–37 W). 
As a result, Versal maintains a consistent advantage in energy efficiency across all configurations, achieving 4.18$\times$ higher efficiency at 30$\times$10 and 2.62$\times$ at 60$\times$20. 
The narrowing gap at larger input sizes reflects the T4 achieving better utilization as the computation grows, while Versal's advantage remains substantial.

\subsection{Impact of Custom Placement}

\begin{figure}[t]
  \centering
  \includegraphics[width=1
  \linewidth]{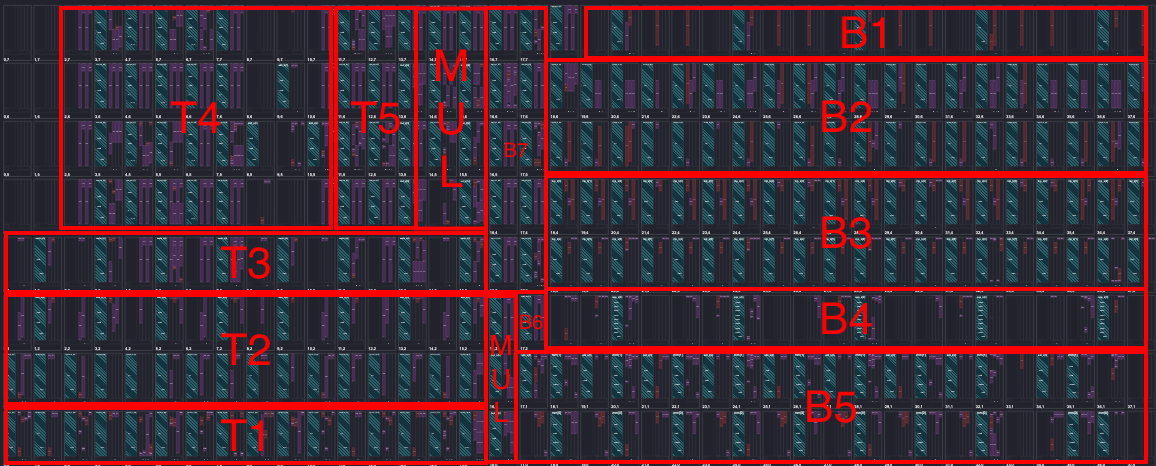}
  \caption{CRONet subgraphs of TrunkNet (T1 to T5) and BranchNet (B1 to B7) mapped physically on the AIE-ML array}
  \label{fig:array_map}
  \vspace{-2mm}
\end{figure}

\begin{table}[t]
\centering\small
\caption{Impact of AIE kernel placement strategy on routing success and compilation time.}
\label{tab:placement_res}
\begin{tabular}{lcc}
\toprule
\textbf{AIE Kernel Placement} &\makecell{\textbf{Time Taken} \\ \textbf{(minutes)}} &\makecell{\textbf{Routing Success} \\ \textbf{(Yes/No)}} \\\midrule
Default compiler settings & 62 &No \\
\makecell[l]{Max compiler \\ optimization level} & 48 &No \\
Custom placement &8 &Yes \\
\bottomrule
\end{tabular}
\vspace{-4mm}
\end{table}

Table \ref{tab:placement_res} summarizes the impact of different placement strategies on compilation outcome and time. 
At the scale of the full CRONet design, the default compiler placement fails to produce a valid routing solution after 62 minutes, as the automatic placer is unable to resolve the routing congestion introduced by 223 active AIE engines and their associated interconnections. 
Increasing the compiler optimization level to its maximum setting does not resolve this issue, with compilation still failing after 48 minutes.
%To overcome this, we develop a custom congestion-aware placement strategy. 
%The approach is based on a straightforward principle. 

Figure \ref{fig:array_map} illustrates the physical mapping of all CRONet layers on the AIE-ML array using our custom placement strategy, where dataflow-adjacent kernels are co-located to minimize routing congestion.
Beyond enabling compilation of large-scale designs that would otherwise fail, this strategy  significantly reduces compilation time, making iterative design exploration practical during development.
As shown in Table \ref{tab:placement_res}, our custom placement achieves successful compilation in 8 minutes at such high AIE utilization. 

\subsection{Discussion}
This work reveals several architectural and programming challenges encountered when targeting the Versal platform, along with potential directions for improvement.

\textbf{Versal architecture constraints:} Versal architectures deliver impressive computational specifications through their AIE-ML arrays, yet they suffer from constrained GMIO bandwidth and limited GMIO ports.
\rev{The limited number of buffer descriptors available in Memory Tiles restricts the complexity of data transformations that can be performed, limiting flexibility in how data is staged and reorganized between layers.}

\red{\textbf{Programming Model Constraints:} The AIE-ML programming model enforces a rigid dataflow paradigm, fundamentally diverging from the flexible control-flow models of GPUs or CPUs. 
Kernels operate exclusively within predefined graphs that execute synchronously for a fixed number of iterations.
While multiple independent graphs enable concurrent execution, the model limits selective iteration counts for subgraphs within a single graph. 
This restriction constrains dataflow benefits, limiting optimization opportunities and increasing development complexity.
Enhancing the programming model to support per-subgraph iteration control would facilitate more complex dataflow designs.
Furthermore, introducing runtime graph reconfiguration could yield substantial performance benefits. 
Currently, resources allocated to a graph are static and cannot be reused by other graphs, leading to underutilization. 
Both these improvements would increase device utilization and further enhance system performance.
}

%\vspace{-3mm}
\section{Conclusion}
\begin{comment}
In this work, we successfully addressed the critical computational bottleneck in Topology Optimization (TO) by accelerating the Convolutional Recurrent Operator Approximator Network (\textbf{CRONet}) on the \textbf{AMD Versal Adaptive Compute Acceleration Platform (ACAP)}. 
We perform end-to-end implementation of CRONet by implementing every layer using a different style of mapping on the AIE-ML engines. 
The performance evaluation demonstrates that the Versal based CRONet \rev{inference achieves up to 2.49$\times$ latency improvements and 4.18$\times$ improvements in energy efficiency over the Nvidia T4 GPU.}
\end{comment}

In this work, we accelerate a complex CNN-RNN hybrid network for Topology Optimization onto the AMD Versal AIE-ML architecture. By designing a fully on-chip, dataflow-oriented implementation that exploits parallelism across the AIE array and minimizes off-chip memory access, we overcome key limitations of conventional GPU-based inference for iterative workloads. Our use of fusion strategies and congestion-aware placement further ensures scalable and efficient execution. Experimental results confirm that this approach delivers significant improvements over an inference-class GPU, highlighting the promise of Versal AIE-ML for real-time, energy-constrained Topology Optimization in digital twin and structural health monitoring applications.

Beyond CRONet, the proposed design methodology demonstrates the broader potential of Versal AIE-ML for deploying complex, irregular neural network workloads that are difficult to efficiently map onto conventional architectures. Our open-source reusable operator library  design enables accelerating other models on the platform. 
Future work will focus on incorporating additional system-level functions like sensor processing on the programmable logic to enable a end-to-end digital twin implementation.

% \section{Introduction}

%% The next two lines define the bibliography style to be used, and
%% the bibliography file.
\bibliographystyle{acm}
% \bibliography{sample-base}
\bibliography{refs}

@INPROCEEDINGS{xu:2025:dt_edge_ai,
  author={Xu, Bin and Banerjee, Ayan and Urooj, Midhat and Gupta, Sandeep K.S.},
  booktitle={2025 IEEE 68th International Midwest Symposium on Circuits and Systems (MWSCAS)}, 
  title={Accelerated Digital Twin Learning for Edge AI: A Comparison of FPGA and Mobile GPU}, 
  year={2025},
  volume={},
  number={},
  pages={149-153},
  doi={10.1109/MWSCAS53549.2025.11244413}}

@article{xu:2025:fast_dt,
  title={Fast Online Digital Twinning on FPGA for Mission Critical Applications},
  author={Xu, Bin and Banerjee, Ayan and Gupta, Sandeep KS},
  journal={arXiv preprint arXiv:2512.17942},
  year={2025}
}

@article{wang:2025:nuclear_dt,
title = {Detailed transient modeling and FPGA-based real-time digital-twin development for sodium-cooled fast reactor},
journal = {Progress in Nuclear Energy},
volume = {189},
pages = {105890},
year = {2025},
issn = {0149-1970},
doi = {https://doi.org/10.1016/j.pnucene.2025.105890},
url = {https://www.sciencedirect.com/science/article/pii/S0149197025002884},
author = {Yulin Wang and Weiran Chen and Venkata Dinavahi}}

@INPROCEEDINGS{nwoke:2023:power_system_converter_dt,
  author={Nwoke, Justus and Milanesi, Marco and Viola, Jairo and Chen, YangQuan},
  booktitle={2023 IEEE 3rd International Conference on Digital Twins and Parallel Intelligence (DTPI)}, 
  title={FPGA-Based Digital Twin Implementation for Power Converter System Monitoring}, 
  year={2023},
  volume={},
  number={},
  pages={1-6},
  doi={10.1109/DTPI59677.2023.10365466}}

@article{deaton:2014:to_review,
	title = {A survey of structural and multidisciplinary continuum topology optimization: post 2000},
	volume = {49},
	issn = {1615-1488},
	url = {https://doi.org/10.1007/s00158-013-0956-z},
	doi = {10.1007/s00158-013-0956-z},
	number = {1},
	journal = {Structural and Multidisciplinary Optimization},
	author = {Deaton, Joshua D. and Grandhi, Ramana V.},
	month = jan,
	year = {2014},
	pages = {1--38},
}

@article{olabiyi:2025:cronet,
title = {CRONet: A convolutional recurrent operator approximator network to accelerate topology optimization},
journal = {Manufacturing Letters},
volume = {44},
pages = {1052-1063},
year = {2025},
note = {53rd SME North American Manufacturing Research Conference (NAMRC 53)},
issn = {2213-8463},
doi = {https://doi.org/10.1016/j.mfglet.2025.06.125},
url = {https://www.sciencedirect.com/science/article/pii/S2213846325001579},
author = {Ridwan Olabiyi and Hui Yang and Ashif Iquebal}
}

@INPROCEEDINGS{hesse:2023:to_fpga,
  author={Hesse, Kasper and Schoeberl, Martin and Aage, Niels and Träff, Erik},
  booktitle={2023 26th Euromicro Conference on Digital System Design (DSD)}, 
  title={On the Feasibility of using FPGA's for Efficient Topology Optimization}, 
  year={2023},
  volume={},
  number={},
  pages={242-250},
  doi={10.1109/DSD60849.2023.00043}}

@INPROCEEDINGS{mhatre:2024:gama,
  author={Kaustubh Mhatre and Endri Taka and Aman Arora},
  booktitle={2025 35th International Conference on Field-Programmable Logic and Applications (FPL25)}, 
  title={GAMA: High-Performance GEMM Acceleration on AMD Versal ML-Optimized AI Engines}, 
  year={2025},
  volume={},
  number={}
}

@article{lu:2021:deeponet,
	title = {Learning nonlinear operators via {DeepONet} based on the universal approximation theorem of operators},
	volume = {3},
	issn = {2522-5839},
	url = {https://doi.org/10.1038/s42256-021-00302-5},
	doi = {10.1038/s42256-021-00302-5},
	number = {3},
	journal = {Nature Machine Intelligence},
	author = {Lu, Lu and Jin, Pengzhan and Pang, Guofei and Zhang, Zhongqiang and Karniadakis, George Em},
	month = mar,
	year = {2021},
	pages = {218--229},
}

@article{catbas:2008:shm_to,
title = {Structural health monitoring and reliability estimation: Long span truss bridge application with environmental monitoring data},
journal = {Engineering Structures},
volume = {30},
number = {9},
pages = {2347-2359},
year = {2008},
issn = {0141-0296},
doi = {https://doi.org/10.1016/j.engstruct.2008.01.013},
url = {https://www.sciencedirect.com/science/article/pii/S014102960800014X},
author = {F. Necati Catbas and Melih Susoy and Dan M. Frangopol}
}

@article{stillmaker:2017:scaling,
title = {Scaling equations for the accurate prediction of CMOS device performance from 180nm to 7nm},
journal = {Integration},
volume = {58},
pages = {74-81},
year = {2017},
issn = {0167-9260},
doi = {https://doi.org/10.1016/j.vlsi.2017.02.002},
url = {https://www.sciencedirect.com/science/article/pii/S0167926017300755},
author = {Aaron Stillmaker and Bevan Baas}
}

@inproceedings{perryman:2023:ac:evaluation,
	title = {{Evaluation of {Xilinx} {Versal} {Architecture} for {Next}-{Gen} {Edge} {Computing} in {Space}}},
	doi = {10.1109/AERO55745.2023.10115906},
	booktitle = {2023 {IEEE} {Aerospace} {Conference}},
	author = {Perryman, Noah and Wilson, Christopher and George, Alan},
	month = mar,
	year = {2023},
	note = {ISSN: 1095-323X},
	pages = {1--11},
}

@inproceedings{yemme:2023:ijcnn:scalable,
	title = {{A {Scalable} {GPT}-2 {Inference} {Hardware} {Architecture} on {FPGA}}},
	doi = {10.1109/IJCNN54540.2023.10191067},
	booktitle = {2023 {International} {Joint} {Conference} on {Neural} {Networks} ({IJCNN})},
	author = {Yemme, Anil and Garani, Shayan Srinivasa},
	month = jun,
	year = {2023},
	note = {ISSN: 2161-4407},
	pages = {1--8},
}

@inproceedings{zhuang:2023:fpga:charm,
	address = {Monterey CA USA},
	title = {{{CHARM}: {C} omposing {H} eterogeneous {A} ccele {R} ators for {M} atrix {Multiply} on {Versal} {ACAP} {Architecture}}},
	isbn = {978-1-4503-9417-8},
	shorttitle = {{CHARM}},
	url = {https://dl.acm.org/doi/10.1145/3543622.3573210},
	doi = {10.1145/3543622.3573210},
	language = {en},
	urldate = {2023-09-21},
	booktitle = {Proceedings of the 2023 {ACM}/{SIGDA} {International} {Symposium} on {Field} {Programmable} {Gate} {Arrays}},
	publisher = {ACM},
	author = {Zhuang, Jinming and Lau, Jason and Ye, Hanchen and Yang, Zhuoping and Du, Yubo and Lo, Jack and Denolf, Kristof and Neuendorffer, Stephen and Jones, Alex and Hu, Jingtong and Chen, Deming and Cong, Jason and Zhou, Peipei},
	month = feb,
	year = {2023},
	pages = {153--164},
}

@misc{zhuang:2023:fpga:automm,
	title = {{{AutoMM}: {Energy}-{Efficient} {Multi}-{Data}-{Type} {Matrix} {Multiply} {Design} on {Heterogeneous} {Programmable} {System}-on-{Chip}}},
	shorttitle = {{AutoMM}},
	url = {http://arxiv.org/abs/2305.18698},
	urldate = {2023-09-26},
	publisher = {arXiv},
	author = {Zhuang, Jinming and Yang, Zhuoping and Zhou, Peipei},
	month = may,
	year = {2023},
	note = {arXiv:2305.18698 [cs]},
}

@inproceedings{chen:2023:fpl:exploiting,
	address = {Gothenburg, Sweden},
	title = {{Exploiting {On}-{Chip} {Heterogeneity} of {Versal} {Architecture} for {GNN} {Inference} {Acceleration}}},
	isbn = {9798350341515},
	url = {https://ieeexplore.ieee.org/document/10296434/},
	doi = {10.1109/FPL60245.2023.00038},
	language = {en},
	urldate = {2023-11-18},
	booktitle = {2023 33rd {International} {Conference} on {Field}-{Programmable} {Logic} and {Applications} ({FPL})},
	publisher = {IEEE},
	author = {Chen, Paul and Manjunath, Pavan and Wijeratne, Sasindu and Zhang, Bingyi and Prasanna, Viktor},
	month = sep,
	year = {2023},
	pages = {219--227},
}

@misc{taka:2023:fpt:maxeva,
	title = {{{MaxEVA}: {Maximizing} the {Efficiency} of {Matrix} {Multiplication} on {Versal} {AI} {Engine}}},
	shorttitle = {{MaxEVA}},
	url = {http://arxiv.org/abs/2311.04980},
	urldate = {2023-12-02},
	publisher = {arXiv},
	author = {Taka, Endri and Arora, Aman and Wu, Kai-Chiang and Marculescu, Diana},
	month = nov,
	year = {2023},
	note = {arXiv:2311.04980 [cs]},
}

@misc{zhuang:2024:fpga:ssr,
	title = {{{SSR}: {Spatial} {Sequential} {Hybrid} {Architecture} for {Latency} {Throughput} {Tradeoff} in {Transformer} {Acceleration}}},
	shorttitle = {{SSR}},
	url = {http://arxiv.org/abs/2401.10417},
	doi = {10.1145/3626202.3637569},
	urldate = {2024-03-09},
	author = {Zhuang, Jinming and Yang, Zhuoping and Ji, Shixin and Huang, Heng and Jones, Alex K. and Hu, Jingtong and Shi, Yiyu and Zhou, Peipei},
	month = feb,
	year = {2024},
	note = {arXiv:2401.10417 [cs]},
}

@inproceedings{yang:2023:iccad:aim,
	address = {San Francisco, CA, USA},
	title = {{{AIM}: {Accelerating} {Arbitrary}-{Precision} {Integer} {Multiplication} on {Heterogeneous} {Reconfigurable} {Computing} {Platform} {Versal} {ACAP}}},
	isbn = {9798350322255},
	shorttitle = {{AIM}},
	url = {https://ieeexplore.ieee.org/document/10323754/},
	doi = {10.1109/ICCAD57390.2023.10323754},
	language = {en},
	urldate = {2024-03-18},
	booktitle = {2023 {IEEE}/{ACM} {International} {Conference} on {Computer} {Aided} {Design} ({ICCAD})},
	publisher = {IEEE},
	author = {Yang, Zhuoping and Zhuang, Jinming and Yin, Jiaqi and Yu, Cunxi and Jones, Alex K. and Zhou, Peipei},
	month = oct,
	year = {2023},
	pages = {1--9},
}

@inproceedings{zhang:2022:fpl:hgcn,
	title = {{H-{GCN}: {A} {Graph} {Convolutional} {Network} {Accelerator} on {Versal} {ACAP} {Architecture}}},
	shorttitle = {H-{GCN}},
	url = {https://ieeexplore.ieee.org/document/10035160},
	doi = {10.1109/FPL57034.2022.00040},
	urldate = {2024-03-18},
	booktitle = {2022 32nd {International} {Conference} on {Field}-{Programmable} {Logic} and {Applications} ({FPL})},
	author = {Zhang, Chengming and Geng, Tong and Guo, Anqi and Tian, Jiannan and Herbordt, Martin and Li, Ang and Tao, Dingwen},
	month = aug,
	year = {2022},
	note = {ISSN: 1946-1488},
	pages = {200--208},
}

@misc{amd:web:versal,
  author = {AMD},
  title = {{AMD Versal ACAP }},
  howpublished = "\url{https://www.amd.com/en/products/adaptive-socs-and-fpgas/versal.html}",
  year = {2024}, 
  note = "[Online; accessed 02-may-2024]"
}

@INPROCEEDINGS{chatarasi:2020:hpec:vyasa,
  author={Chatarasi, Prasanth and Neuendorffer, Stephen and Bayliss, Samuel and Vissers, Kees and Sarkar, Vivek},
  booktitle={2020 IEEE High Performance Extreme Computing Conference (HPEC)}, 
  title={{Vyasa: A High-Performance Vectorizing Compiler for Tensor Convolutions on the Xilinx AI Engine}}, 
  year={2020},
  volume={},
  number={},
  pages={1-10},
  doi={10.1109/HPEC43674.2020.9286183}}

@misc{amd:2022:web:api_user_guide,
  title = {{AI Engine API User Guide.}},
  year = {2022}
}

@misc{amd:2023:web:aie2_arch,
  author = {AMD/Xilinx},
  title = {{
Versal Adaptive SoC AIE-ML Architecture Manual (AM020)}},
  year = {2023}
}

@misc{AMD:2025:git:iron_tollkit,
  author = {AMD},
  title = {{IRON: Unlocking the Full Potential of NPUs}},
url = {
https://github.com/amd/IRON/tree/devel},
  year = {2025}
}

@misc{amd:2023:web:sc,
  author = {AMD/Xilinx},
  title = {{
VEK280 Evaluation Board User Guide (UG1612)}},
  year = {2023}
}

@inproceedings{singh:2023:ics:sparta,
  title={{SPARTA: Spatial Acceleration for Efficient and Scalable Horizontal Diffusion Weather Stencil Computation}},
  author={Singh, Gagandeep and Khodamoradi, Alireza and Denolf, Kristof and Lo, Jack and G{\'o}mez-Luna, Juan and Melber, Joseph and Bisca, Andra and Corporaal, Henk and Mutlu, Onur},
  booktitle={ICS},
  year={2023}
}

@MASTERSTHESIS{wierse:2023:ethz:evaluation,
	copyright = {In Copyright - Non-Commercial Use Permitted},
	year = {2023-02},
	type = {Bachelor Thesis},
	author = {Wierse, Max},
	size = {62 p.},
	language = {en},
	address = {Zurich},
	publisher = {ETH Zurich},
	DOI = {10.3929/ethz-b-000600880},
	title = {{Evaluation of Xilinx Versal Device}},
	school = {ETH Zurich}
}

@article{zhuang:2024:trets:charm_2,
author = {Zhuang, Jinming and Lau, Jason and Ye, Hanchen and Yang, Zhuoping and Ji, Shixin and Lo, Jack and Denolf, Kristof and Neuendorffer, Stephen and Jones, Alex and Hu, Jingtong and Shi, Yiyu and Chen, Deming and Cong, Jason and Zhou, Peipei},
title = {{CHARM 2.0: Composing Heterogeneous Accelerators for Deep Learning on Versal ACAP Architecture}},
year = {2024},
issue_date = {September 2024},
publisher = {Association for Computing Machinery},
address = {New York, NY, USA},
volume = {17},
number = {3},
issn = {1936-7406},
url = {https://doi.org/10.1145/3686163},
doi = {10.1145/3686163},
journal = {ACM Trans. Reconfigurable Technol. Syst.},
month = sep,
articleno = {51},
numpages = {31}
}

@inproceedings{zhuang:2025:fpga:aries,
author = {Zhuang, Jinming and Xiang, Shaojie and Chen, Hongzheng and Zhang, Niansong and Yang, Zhuoping and Mao, Tony and Zhang, Zhiru and Zhou, Peipei},
title = {{ARIES: An Agile MLIR-Based Compilation Flow for Reconfigurable Devices with AI Engines}},
year = {2025},
isbn = {9798400713965},
publisher = {Association for Computing Machinery},
address = {New York, NY, USA},
url = {https://doi.org/10.1145/3706628.3708870},
doi = {10.1145/3706628.3708870},
booktitle = {Proceedings of the 2025 ACM/SIGDA International Symposium on Field Programmable Gate Arrays},
pages = {92–102},
numpages = {11},
location = {Monterey, CA, USA},
series = {FPGA '25}
}

@INPROCEEDINGS{deng:2024:fpl:ama,
  author={Deng, Xiaodong and Wang, Shijie and Gao, Tianyi and Liu, Jing and Liu, Longjun and Zheng, Nanning},
  booktitle={2024 34th International Conference on Field-Programmable Logic and Applications (FPL)}, 
  title={{AMA: An Analytical Approach to Maximizing the Efficiency of Deep Learning on Versal AI Engine}}, 
  year={2024},
  volume={},
  number={},
  pages={227-235},
  doi={10.1109/FPL64840.2024.00039}}

@ARTICLE{dong:2024:tcad:eq_vit,
  author={Dong, Peiyan and Zhuang, Jinming and Yang, Zhuoping and Ji, Shixin and Li, Yanyu and Xu, Dongkuan and Huang, Heng and Hu, Jingtong and Jones, Alex K. and Shi, Yiyu and Wang, Yanzhi and Zhou, Peipei},
  journal={IEEE Transactions on Computer-Aided Design of Integrated Circuits and Systems}, 
  title={{EQ-ViT: Algorithm-Hardware Co-Design for End-to-End Acceleration of Real-Time Vision Transformer Inference on Versal ACAP Architecture}}, 
  year={2024},
  volume={43},
  number={11},
  pages={3949-3960},
  doi={10.1109/TCAD.2024.3443692}}

@misc{wang:2025:arxiv:rsn_xnn,
      title={{Reconfigurable Stream Network Architecture}}, 
      author={Chengyue Wang and Xiaofan Zhang and Jason Cong and James C. Hoe},
      year={2025},
      eprint={2411.17966},
      archivePrefix={arXiv},
      primaryClass={cs.AR},
      url={https://arxiv.org/abs/2411.17966}, 
}

@misc{lattner:2020:axriv:mlir,
      title={{MLIR: A Compiler Infrastructure for the End of Moore's Law}}, 
      author={Chris Lattner and Mehdi Amini and Uday Bondhugula and Albert Cohen and Andy Davis and Jacques Pienaar and River Riddle and Tatiana Shpeisman and Nicolas Vasilache and Oleksandr Zinenko},
      year={2020},
      eprint={2002.11054},
      archivePrefix={arXiv},
      primaryClass={cs.PL},
      url={https://arxiv.org/abs/2002.11054}, 
}

@INPROCEEDINGS{mhatre:2025:ispass:performance,
  author={Mhatre, Kaustubh Manohar and Mulleti, Venkata Guru Prashanth and Bansil, Curt John and Taka, Endri and Arora, Aman},
  booktitle={2025 IEEE International Symposium on Performance Analysis of Systems and Software (ISPASS)}, 
  title={Performance Analysis of GEMM Workloads on the AMD Versal Platform}, 
  year={2025},
  volume={},
  number={},
  pages={150-161},
  doi={10.1109/ISPASS64960.2025.00023}
}

@inproceedings{elliott:2025:mm_aie_cpu,
author = {Binder, Elliott D. and Low, Jeffrey and Low, Tze Meng},
title = {Architecture-Aware Models of AI Engines for High-Performance Matrix Matrix Multiplication},
year = {2025},
isbn = {9798400720741},
publisher = {Association for Computing Machinery},
address = {New York, NY, USA},
url = {https://doi.org/10.1145/3754598.3754612},
doi = {10.1145/3754598.3754612},
booktitle = {Proceedings of the 54th International Conference on Parallel Processing},
pages = {531–540},
numpages = {10},
location = {
},
series = {ICPP '25}
}

%%
%% If your work has an appendix, this is the place to put it.
% \appendix

% \section{Research Methods}

% \subsection{Part One}

% Lorem ipsum dolor sit amet, consectetur adipiscing elit. Morbi
% malesuada, quam in pulvinar varius, metus nunc fermentum urna, id
% sollicitudin purus odio sit amet enim. Aliquam ullamcorper eu ipsum
% vel mollis. Curabitur quis dictum nisl. Phasellus vel semper risus, et
% lacinia dolor. Integer ultricies commodo sem nec semper.

% \subsection{Part Two}

% Etiam commodo feugiat nisl pulvinar pellentesque. Etiam auctor sodales
% ligula, non varius nibh pulvinar semper. Suspendisse nec lectus non
% ipsum convallis congue hendrerit vitae sapien. Donec at laoreet
% eros. Vivamus non purus placerat, scelerisque diam eu, cursus
% ante. Etiam aliquam tortor auctor efficitur mattis.

% \section{Online Resources}

% Nam id fermentum dui. Suspendisse sagittis tortor a nulla mollis, in
% pulvinar ex pretium. Sed interdum orci quis metus euismod, et sagittis
% enim maximus. Vestibulum gravida massa ut felis suscipit
% congue. Quisque mattis elit a risus ultrices commodo venenatis eget
% dui. Etiam sagittis eleifend elementum.

% Nam interdum magna at lectus dignissim, ac dignissim lorem
% rhoncus. Maecenas eu arcu ac neque placerat aliquam. Nunc pulvinar
% massa et mattis lacinia.

\end{document}